# A new development of modified quantum mechanics

Yu-kuo Zhao[1,†], and Yu-xin Dong[2]

[1]Department of Mechanical Engineering, Shanghai Jiao Tong University, Shanghai, 200240, China

[2]School of Digital Media & Design Arts, Beijing University of Posts and Telecommunications, Beijing, 100876, China

**Abstract**

Here, we propose a new modified quantum mechanics and its new algorithms of atomic fine-structure—asymmetric variational method based on hydrogen-like atom orbit. In addition, as we all know, the ab initio calculation of atomic fine-structure is the first driving force to create relativistic quantum mechanics, so every new development of it is an interesting and important hot topic. But we will not try to prove (or refute) strictly, because there is no analytical solution to the multi-body Schrödinger equation and the magnetic potential integral of ampere force of Langevin, so we use our method to calculate the atom fine-structure of within the second period, and prove that our calculation is closer to the experimental observation value through this numerical experiment method (Our calculation accuracy is better than those of relativistic Hartree-Fock method). Therefore, our discussion is a positive exploration and conjecture.

## I. Introduction

The moment quantum mechanics was founded[1-4], physicists knew that the energy level structure of Schrödinger equation under Coulomb action potential will not be consistent with the experiment, because in the Bohr-Sommerfeld model[5,6] in 1916, the energy level formula of its hydrogen-like atom is closer to the experimental observation value, not Bohr-Schrödinger.

But nevertheless, the Schrödinger equation is still an indispensable first principle, because the multi-body Dirac equation is difficult to construct and calculate[7,8]—In the modern computer program of quantum mechanics and quantum chemistry, it is common and necessary steps to firstly calculate the approximate solution of the Schrödinger equation under the Coulomb action potential without time, and then use its wave function to correct its energy level, such as the famous Gaussian-16 program[9]. So, we thought about a question, is there an eigenequation similar to Schrödinger, which can combine the above two calculation steps into one?

The answer is yes, but some the relativistic quantum electrodynamics effect needs to be approximated by a fitted magnetic potential[10], and interpreted it as the interaction of two loop currents.

In Langevin opinion, the circular motion of electrons and nuclei, like two loop circuits without resistance, will produce an induced magnetic field, namely ampere

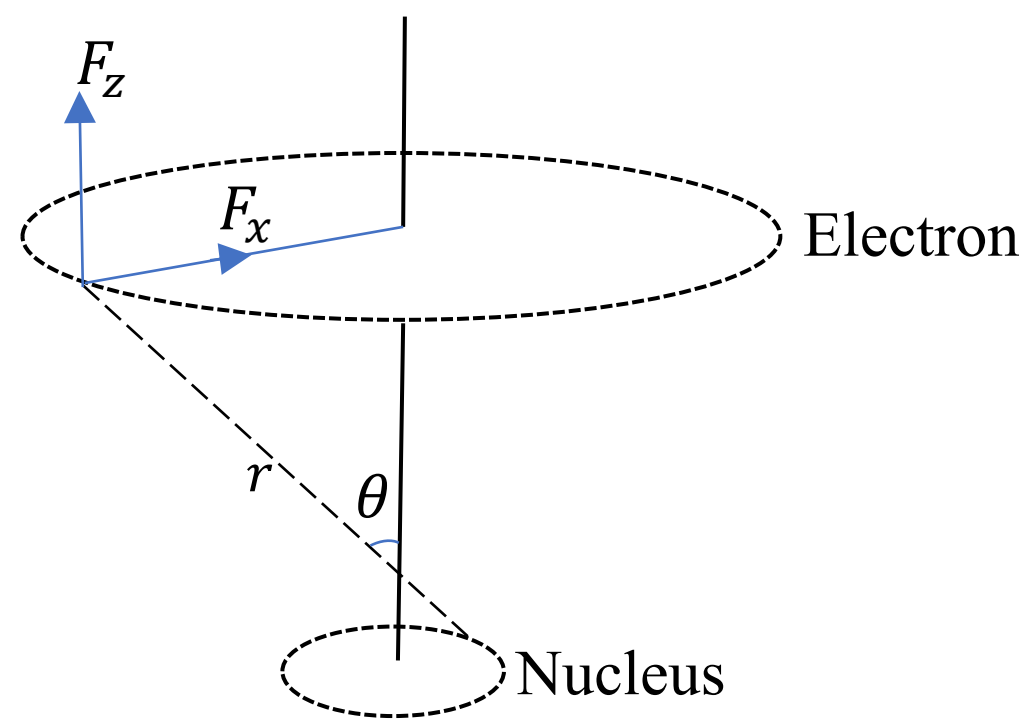

Fig. 1. Ampere force in the movement of electrons around the nucleus. Where, the component in the horizontal direction is defined to be $F_x$, the component in the vertical direction is defined to be $F_z$.

force[11], as shown in Fig.1.

In Fig.1, according to Biot-Savart law in magnetostatics, our ampere force is

$$\begin{cases} F_x = \dfrac{1}{4\pi\varepsilon_0}\dfrac{Ze^2}{r^2}\dfrac{m_{\mathrm{n}}\sin^2(\theta)f_x}{m_{\mathrm{e}}(m_{\mathrm{n}}+m_{\mathrm{e}})r-m_{\mathrm{n}}\sin(\theta)f_x} \\ F_z = \dfrac{1}{4\pi\varepsilon_0}\dfrac{Ze^2}{r^2}\dfrac{m_{\mathrm{n}}\sin^2(\theta)f_z}{m_{\mathrm{e}}(m_{\mathrm{n}}+m_{\mathrm{e}})r-m_{\mathrm{n}}\sin(\theta)f_x} \end{cases} \quad (1).$$

where, the mass of electron is defined to be $m_{\mathrm{e}}$, the mass of nucleus is defined to be $m_{\mathrm{n}}$, the nuclear charge number of nucleus is defined to be $Z$, the charge on an electron is defined to be $e$, the magnetoconductivity is defined to be $\mu_0$, the dielectric coefficient is defined to be $\varepsilon_0$, the intermediate substitution function is defined to be

$$\begin{aligned} f_x = {} & \mu_0\frac{Ze^2}{16\pi^2}\frac{(m_{\mathrm{n}}+m_{\mathrm{e}})}{m_{\mathrm{n}}}\frac{1}{\sin(\theta)}+\mu_0\frac{Ze^2}{16\pi^2}\frac{m_{\mathrm{e}}}{(m_{\mathrm{n}}+m_{\mathrm{e}})}\sin(\theta) \\ & -\mu_0\frac{Ze^2}{16\pi^2}\frac{\left(m_{\mathrm{n}}^2+m_{\mathrm{e}}^2\cos(2\theta)+2m_{\mathrm{n}}m_{\mathrm{e}}\cos^2(\theta)\right)(m_{\mathrm{n}}+m_{\mathrm{e}})}{\left(m_{\mathrm{n}}^2+m_{\mathrm{e}}^2+2m_{\mathrm{n}}m_{\mathrm{e}}\cos(2\theta)\right)m_{\mathrm{n}}}\frac{1}{\sin(\theta)} \\ & +\mu_0\frac{Ze^2}{16\pi^2}\frac{\left(m_{\mathrm{n}}^2+m_{\mathrm{e}}^2\cos(2\theta)+2m_{\mathrm{n}}m_{\mathrm{e}}\cos^2(\theta)\right)m_{\mathrm{e}}}{\left(m_{\mathrm{n}}^2+m_{\mathrm{e}}^2+2m_{\mathrm{n}}m_{\mathrm{e}}\cos(2\theta)\right)(m_{\mathrm{n}}+m_{\mathrm{e}})}\sin(\theta) \end{aligned} \quad (2).$$

$$\begin{aligned} f_z = {} & \mu_0\frac{Ze^2}{8\pi}\frac{\left(m_{\mathrm{n}}^2+m_{\mathrm{e}}^2+2m_{\mathrm{n}}m_{\mathrm{e}}\cos^2(\theta)\right)(m_{\mathrm{n}}+m_{\mathrm{e}})^2}{\left(m_{\mathrm{n}}^2+m_{\mathrm{e}}^2+2m_{\mathrm{n}}m_{\mathrm{e}}\cos(2\theta)\right)m_{\mathrm{n}}^2}\frac{\cos(\theta)}{\sin^2(\theta)}-\mu_0\frac{Ze^2}{8\pi}\frac{m_{\mathrm{e}}}{m_{\mathrm{n}}}\cos(\theta) \\ & -\mu_0\frac{Ze^2}{8\pi}\frac{\left(m_{\mathrm{n}}^2+m_{\mathrm{e}}^2+2m_{\mathrm{n}}m_{\mathrm{e}}\cos^2(\theta)\right)m_{\mathrm{e}}}{\left(m_{\mathrm{n}}^2+m_{\mathrm{e}}^2+2m_{\mathrm{n}}m_{\mathrm{e}}\cos(2\theta)\right)m_{\mathrm{n}}}\cos(\theta)-\mu_0\frac{Ze^2}{8\pi}\frac{(m_{\mathrm{n}}+m_{\mathrm{e}})^2}{m_{\mathrm{n}}^2}\frac{\cos(\theta)}{\sin^2(\theta)} \end{aligned} \quad (3).$$

But in fact, because it is difficult to get an analytical solution for the magnetic potential integral of $\mathrm{V_M}(r)=\int_r^{+\infty}(\sin(\theta)F_x-\cos(\theta)F_z)dr$, we will not strictly prove (or refute) it. And the real motion may also need to consider other factors, such as the elliptical trajectory of particles, the acceleration in the vertical direction and the recoil caused by thermal radiation—Therefore, we shelve general derivation of magnetic potential here, but numerical experiments can support that our answer is an interesting and credible conclusion—In the numerical experiment of hydrogen-like atoms, our maximum error rate comes from $3D_{5/2} - 3D_{3/2}$, is 13.4%, but other energy level intervals are consistent with the fine structure in quantum electrodynamics[12], and our wave function is about equal to the classical solution of Schrödinger equation under Coulomb action potential, so it is also suitable for the theoretical calculation of ultra-fine intervals in quantum electrodynamics[12,13,14] without further explanation.

Then, we need to think about how to interpret the multiplicity of atoms.

In the non-relativistic Hartree-Fock method (Self-consistent field theory without spin function), there is no singlet solution. Therefore, the introduction of spin function into quantum mechanics has become the consensus of many physicists, and it is considered as a key and unique concept in relativistic quantum mechanics[8,15-29].

However, spin is ghostly, because there is no explicit wave function to describe its properties—The independent variable of spin wave function is ambiguous[30], so all refs avoid its explicit expression, which is rare in physics. Therefore, a brief review of its early history seems to make us understand it better.

In 1921, Compton discussed the magnetism of atoms and was the first to put forward the suggestion of quantum spin of electrons—The electron itself has a tiny and incessant rotating motion like a gyro observer[11].

In 1922, Stern and Gerlach confirmed the quantization of atomic angular momentum—during this period, the Bohr model holds that the alkali metal beam will undergo double splitting after passing through a non-uniform magnetic field[31].

In 1925, Uhlenbeck and Goudsmit used the old electron spin theory to explain the mechanism of Zeeman effect and Stern-Gerlach experiment[32,33]—Their theory is similar to Compton's gyro observer, but the volume of electrons obviously does not support this theory.

In 1928, Dirac equation redefined spin as the intrinsic property of particles[34,35]—1/2 spin[36]—but there was no macroscopic motion corresponding to it, and Stern-Gerlach experiment would be redefined as favorable evidence for this new quantum spin theory.

But its application originated from Bohm's EPRB paradox in 1952 —He suggested that experimenters use spin to test the completeness of quantum mechanics[37,38]. And in 1957, Wu and her colleagues used the spin of $^{60}Co$ to test the parity non-conservation law in weak interaction[39]. In 1972, Clauser and Freedman used the linear polarization correlation of photons emitted by the cascade of calcium atoms to test the authenticity of the local hidden variable theory[40], and more experiments and applications can be found in the refs[22-29], omitted.

Therefore, electron spin has a wide range of experimental support, thought it is ignored that its integral needs settings other than mathematics. However, *every cloud has a silver lining*. History suggests that when many physicists have not found a explicit wave function to describe electron spin, it is a positive exploration and conjecture for us to look back at the possibility of Langevin's hypothesis of circular circuit and Schrödinger-like equation. Therefore, we propose an asymmetric variational method (without spin function) in the orbit of hydrogen-like atoms, which is used to calculate the multiple of atoms, and is proved to be consistent with the experimental measurement through the numerical experiment from helium atom to neon atom. That is to say, in the absence of spin wave function, singlet state (the missing root) in Hartree-Fock method is only a mathematical question, which cannot be a natural conclusion of the existence of spin function. But our discussion does not involve experiments beyond the atomic level structure, because this work seems to be more critical and basic—With the Born-Oppenheimer approximation, the following Schrödinger-like equation of atoms can be consistent with all observations of the fine structure of atoms, including those Lamb shift calculations[12,41] in quantum electrodynamics, namely (atomic units):

$$\mathbb{i}\frac{\partial}{\partial t}\Psi = \hat{H}\Psi = \sum_{i=1}^{N}\left(-\frac{1}{2}\nabla_i^2 - \frac{Z}{r_i} + V_{\mathrm{M}}(\vec{\boldsymbol{r}}_i) + \sum_{j=i+1}^{N}\frac{1}{r_{i,j}}\right)\Psi = E\Psi \qquad (4).$$

and Slater-Gaunt integration method[42] can be replaced by our algorithm, because our repulsive energy integration is an analytical solution.

Where, the Laplacian operator is defined to be $\nabla_i^2$, the atomic nuclear charge is defined to be $Z$, the distance between the electron and the nucleus is defined to be $r_i$, the magnetic potential is defined to be $V_{\mathrm{M}}(\vec{\boldsymbol{r}}_i)$, the position of the electron is defined to be $\vec{\boldsymbol{r}}_i$, the distance between electrons is defined to be $r_{i,j}$, the wave function is defined to be $\Psi$, and the energy level is defined to be $E$.

## II. Magnetic potential of hydrogen-like atoms

Let the magnetic potential of hydrogen-like atoms be (the contribution of ampere force generated by the movement of electrons around the nucleus)

$$V_{\mathrm{M}}(\vec{\boldsymbol{r}}) = -\frac{\delta_0(Z)}{r} - \frac{\delta_1(Z)}{r^2\sin^2(\theta)} - \frac{\delta_2(Z)}{r^2\cos^2(\theta)} - \frac{\delta_3(Z)}{r^2} \qquad (5).$$

then the fitting equation of its corresponding action quantity is

$$\begin{cases} \dfrac{\left(Z+\delta_0(Z)\right)^2}{2(0.5+\Lambda_1)^2} = \dfrac{1}{\alpha^2}\left(1-\sqrt{1-\alpha^2 Z^2}+\dfrac{\alpha^2}{4}\Delta_{\mathrm{I}}E(Z)\right) \\ \dfrac{\left(Z+\delta_0(Z)\right)^2}{2(1.5-\Lambda_1)^2} = \dfrac{1}{\alpha^2}\left(1-\sqrt{1-\alpha^2 Z^2}-\dfrac{3\alpha^2}{4}\Delta_{\mathrm{I}}E(Z)\right) \\ \dfrac{\left(Z+\delta_0(Z)\right)^2}{2(1.5+\Lambda_1)^2} - \dfrac{\left(Z+\delta_0(Z)\right)^2}{2(1.5+\Lambda_2)^2} = \Delta_{\mathrm{II}}E(Z) \\ \dfrac{\left(Z+\delta_0(Z)\right)^2}{(1.5+\Lambda_1)^2} - \dfrac{\left(Z+\delta_0(Z)\right)^2}{(0.5+\Lambda_3)^2} = \dfrac{1}{\alpha^2}\left(\sqrt{4-\alpha^2 Z^2}-\sqrt{2+2\sqrt{1-\alpha^2 Z^2}}\right) \end{cases} \quad (6.1).$$

$$\begin{cases} \left(1-\sqrt{0.25-2\delta_2(Z)}-\sqrt{-2\delta_1(Z)}\right)^2 - 2\delta_3(Z) = \Lambda_1^2 \\ \left(1-\sqrt{0.25-2\delta_2(Z)}+\sqrt{-2\delta_1(Z)}\right)^2 - 2\delta_3(Z) = \Lambda_2^2 \\ \left(2-\sqrt{0.25-2\delta_2(Z)}-\sqrt{-2\delta_1(Z)}\right)^2 - 2\delta_3(Z) = \Lambda_3^2 \end{cases} \quad (6.2).$$

and its solution is (as shown in the Table I)

$$\delta_0(Z) = \frac{\sqrt{\frac{8}{\alpha^2}\left(1-\sqrt{1-\alpha^2 Z^2}+\frac{\alpha^2}{4}\Delta_{\mathrm{I}}E(Z)\right)\left(1-\sqrt{1-\alpha^2 Z^2}-\frac{3\alpha^2}{4}\Delta_{\mathrm{I}}E(Z)\right)}}{\sqrt{1-\sqrt{1-\alpha^2 Z^2}+\frac{\alpha^2}{4}\Delta_{\mathrm{I}}E(Z)}+\sqrt{1-\sqrt{1-\alpha^2 Z^2}-\frac{3\alpha^2}{4}\Delta_{\mathrm{I}}E(Z)}} - \mathrm{Z} \quad (7.1).$$

$$\begin{cases} \Lambda_1 = \dfrac{\left(Z+\delta_0(Z)\right)\alpha}{\sqrt{2-2\sqrt{1-\alpha^2 Z^2}+\frac{\alpha^2}{2}\Delta_{\mathrm{I}}E(Z)}} - 0.5 \\ \Lambda_2 = \dfrac{\left(Z+\delta_0(Z)\right)(1.5+\Lambda_1)}{\sqrt{\left(Z+\delta_0(Z)\right)^2 - 2(1.5+\Lambda_1)^2\Delta_{\mathrm{II}}E(Z)}} - 1.5 \\ \Lambda_3 = \dfrac{\left(Z+\delta_0(Z)\right)(1.5+\Lambda_1)\alpha}{\sqrt{\left(Z+\delta_0(Z)\right)^2\alpha^2 - (1.5+\Lambda_1)^2\left(\sqrt{4-\alpha^2 Z^2}-\sqrt{2+2\sqrt{1-\alpha^2 Z^2}}\right)}} - 0.5 \end{cases} \quad (7.2).$$

$$\begin{cases} \delta_1(Z) = -\dfrac{1}{32}\left(1+\Lambda_1^2-\Lambda_3^2+\sqrt{(1+\Lambda_1^2-\Lambda_3^2)^2+4(\Lambda_2^2-\Lambda_1^2)}\right)^2 \\ \delta_2(Z) = 0.125+\dfrac{1}{64\delta_1(Z)}\left(4\sqrt{-2\delta_1(Z)}-\Lambda_2^2+\Lambda_1^2\right)^2 \\ \delta_3(Z) = 0.5\left(1-\sqrt{0.25-2\delta_2(Z)}-\sqrt{-2\delta_1(Z)}\right)^2 - 0.5\Lambda_1^2 \end{cases} \quad (7.3).$$

where, the spherical coordinates of electrons as defined to be $\begin{cases} r = \sqrt{x^2+y^2+z^2} \\ \theta = \mathrm{Arccos}\left(\frac{z}{r}\right) \\ \phi = \mathrm{Arccos}\left(\frac{x}{r\sin(\theta)}\right) \end{cases}$, the fine structure constant is defined to be $\alpha \approx 1/137.036$, the Lamb shift of ground state is defined to be $\Delta_{\mathrm{I}}E(Z)$, the Lamb shift of $(2\mathrm{S}_{1/2}-2\mathrm{P}_{1/2})$ state is defined to be $\Delta_{\mathrm{II}}E(Z)$, and as shown in the reference[41] , omitted. And the unit in this paper is the atomic unit system, such as the reduced Planck constant is $\hbar=1$.

Table I. Action quantity in magnetic potential. Where, the nuclear root-mean-square radius in fermi is defined to be $\langle\dot{r}^2\rangle^{1/2}$, the nuclear equivalent radius in fermi is defined to be $\dot{r}$, and the nuclear mass in a. m. u. is defined to be $m_{\mathrm{n}}$.

| $Z$ | $m_{\mathrm{n}}$ | $\langle\dot{r}^2\rangle^{1/2}$ | $\dot{r}$ | $\delta_0(Z)$ | $\delta_1(Z)$ | $\delta_2(Z)$ | $\delta_3(Z)$ |
|---|---|---|---|---|---|---|---|
| 1 | 1.00728 | 0.809 | 1.044 | 6.346E-06 | -2.0678E-13 | 9.9958E-06 | 9.9852E-06 |
| 2 | 4.00151 | 1.673 | 2.160 | 5.121E-05 | -2.2764E-12 | 3.9985E-05 | 3.9947E-05 |
| 3 | 7.01436 | 2.392 | 3.088 | 1.737E-04 | -8.9666E-12 | 8.9971E-05 | 8.9904E-05 |
| 4 | 9.00999 | 2.519 | 3.252 | 4.132E-04 | -2.3294E-11 | 1.5996E-04 | 1.5989E-04 |
| 5 | 11.0066 | 2.397 | 3.095 | 8.093E-04 | -4.8266E-11 | 2.4997E-04 | 2.4995E-04 |
| 6 | 11.9967 | 2.455 | 3.169 | 1.402E-03 | -8.6858E-11 | 3.6001E-04 | 3.6014E-04 |
| 7 | 13.9992 | 2.549 | 3.291 | 2.230E-03 | -1.4190E-10 | 4.9010E-04 | 4.9053E-04 |
| 8 | 15.9905 | 2.711 | 3.500 | 3.335E-03 | -2.1609E-10 | 6.4026E-04 | 6.4121E-04 |
| 9 | 18.9935 | 2.900 | 3.744 | 4.757E-03 | -3.1207E-10 | 8.1052E-04 | 8.1227E-04 |
| 10 | 19.987 | 3.024 | 3.904 | 6.535E-03 | -4.3180E-10 | 1.0009E-03 | 1.0038E-03 |
| 11 | 22.984 | 2.963 | 3.825 | 8.711E-03 | -5.7758E-10 | 1.2114E-03 | 1.2160E-03 |
| 12 | 23.978 | 3.054 | 3.943 | 1.133E-02 | -7.5216E-10 | 1.4422E-03 | 1.4490E-03 |
| 13 | 26.974 | 3.041 | 3.926 | 1.442E-02 | -9.5589E-10 | 1.6931E-03 | 1.7028E-03 |
| 14 | 27.969 | 3.107 | 4.011 | 1.803E-02 | -1.1921E-09 | 1.9643E-03 | 1.9778E-03 |
| 15 | 30.966 | 3.197 | 4.127 | 2.221E-02 | -1.4620E-09 | 2.2559E-03 | 2.2740E-03 |
| 16 | 31.963 | 3.247 | 4.192 | 2.699E-02 | -1.7667E-09 | 2.5678E-03 | 2.5916E-03 |
| 17 | 34.960 | 3.335 | 4.305 | 3.242E-02 | -2.1081E-09 | 2.9001E-03 | 2.9310E-03 |
| 18 | 39.953 | 3.428 | 4.426 | 3.853E-02 | -2.4873E-09 | 3.2529E-03 | 3.2923E-03 |
| 19 | 38.953 | 3.407 | 4.398 | 4.538E-02 | -2.9044E-09 | 3.6262E-03 | 3.6757E-03 |
| 20 | 39.952 | 3.476 | 4.487 | 5.300E-02 | -3.3601E-09 | 4.0201E-03 | 4.0815E-03 |
| 21 | 44.944 | 3.542 | 4.573 | 6.145E-02 | -3.8610E-09 | 4.4347E-03 | 4.5100E-03 |
| 22 | 47.936 | 3.599 | 4.646 | 7.075E-02 | -4.3983E-09 | 4.8699E-03 | 4.9615E-03 |

| 23 | 50.931 | 3.602 | 4.650 | 8.096E-02 | -4.9804E-09 | 5.3260E-03 | 5.4363E-03 |
|---|---|---|---|---|---|---|---|
| 24 | 51.927 | 3.612 | 4.663 | 9.212E-02 | -5.6019E-09 | 5.8030E-03 | 5.9347E-03 |
| 25 | 54.927 | 3.705 | 4.783 | 0.1043 | -6.2733E-09 | 6.3009E-03 | 6.4571E-03 |
| 26 | 55.921 | 3.736 | 4.823 | 0.1175 | -6.9855E-09 | 6.8198E-03 | 7.0039E-03 |
| 27 | 58.918 | 3.782 | 4.883 | 0.1318 | -7.7422E-09 | 7.3599E-03 | 7.5754E-03 |
| 28 | 57.920 | 3.776 | 4.875 | 0.1472 | -8.5406E-09 | 7.9211E-03 | 8.1720E-03 |
| 29 | 62.914 | 3.898 | 5.032 | 0.1638 | -9.3978E-09 | 8.5037E-03 | 8.7943E-03 |
| 30 | 63.913 | 3.955 | 5.106 | 0.1816 | -1.0296E-08 | 9.1076E-03 | 9.4425E-03 |
| 31 | 68.909 | 3.998 | 5.161 | 0.2008 | -1.1237E-08 | 9.7331E-03 | 1.0117E-02 |
| 32 | 73.904 | 4.079 | 5.266 | 0.2212 | -1.2234E-08 | 1.0380E-02 | 1.0819E-02 |
| 33 | 74.903 | 4.104 | 5.298 | 0.2430 | -1.3267E-08 | 1.1049E-02 | 1.1548E-02 |
| 34 | 79.898 | 4.171 | 5.385 | 0.2663 | -1.4354E-08 | 1.1739E-02 | 1.2306E-02 |
| 35 | 78.899 | 4.156 | 5.365 | 0.2911 | -1.5486E-08 | 1.2452E-02 | 1.3092E-02 |
| 36 | 83.892 | 4.230 | 5.461 | 0.3173 | -1.6672E-08 | 1.3186E-02 | 1.3907E-02 |
| 37 | 84.892 | 4.245 | 5.480 | 0.3452 | -1.7880E-08 | 1.3943E-02 | 1.4752E-02 |
| 38 | 87.885 | 4.242 | 5.476 | 0.3747 | -1.9146E-08 | 1.4721E-02 | 1.5628E-02 |
| 39 | 88.884 | 4.244 | 5.479 | 0.4059 | -2.0463E-08 | 1.5523E-02 | 1.6535E-02 |
| 40 | 89.883 | 4.273 | 5.516 | 0.4389 | -2.1819E-08 | 1.6346E-02 | 1.7474E-02 |
| 41 | 92.884 | 4.318 | 5.575 | 0.4737 | -2.3250E-08 | 1.7192E-02 | 1.8446E-02 |
| 42 | 97.882 | 4.415 | 5.700 | 0.5103 | -2.4764E-08 | 1.8061E-02 | 1.9452E-02 |
| 43 | 96.883 | 4.410 | 5.693 | 0.5489 | -2.6247E-08 | 1.8953E-02 | 2.0491E-02 |
| 44 | 101.880 | 4.475 | 5.777 | 0.5895 | -2.7856E-08 | 1.9868E-02 | 2.1566E-02 |
| 45 | 102.880 | 4.502 | 5.812 | 0.6321 | -2.9477E-08 | 2.0806E-02 | 2.2677E-02 |
| 46 | 105.880 | 4.526 | 5.843 | 0.6768 | -3.1140E-08 | 2.1767E-02 | 2.3826E-02 |
| 47 | 106.880 | 4.542 | 5.864 | 0.7237 | -3.2855E-08 | 2.2751E-02 | 2.5012E-02 |
| 48 | 113.880 | 4.613 | 5.955 | 0.7729 | -3.4704E-08 | 2.3759E-02 | 2.6238E-02 |
| 49 | 114.880 | 4.619 | 5.963 | 0.8243 | -3.6508E-08 | 2.4791E-02 | 2.7504E-02 |
| 50 | 119.870 | 4.655 | 6.010 | 0.8782 | -3.8421E-08 | 2.5847E-02 | 2.8811E-02 |
| 51 | 120.880 | 4.704 | 6.073 | 0.9345 | -4.0411E-08 | 2.6926E-02 | 3.0161E-02 |
| 52 | 129.880 | 4.804 | 6.202 | 0.9933 | -4.2568E-08 | 2.8029E-02 | 3.1555E-02 |
| 53 | 126.880 | 4.752 | 6.135 | 1.0547 | -4.4491E-08 | 2.9157E-02 | 3.2994E-02 |
| 54 | 131.870 | 4.826 | 6.230 | 1.1188 | -4.6754E-08 | 3.0308E-02 | 3.4480E-02 |
| 55 | 132.880 | 4.807 | 6.206 | 1.1856 | -4.8867E-08 | 3.1485E-02 | 3.6013E-02 |
| 56 | 137.870 | 4.840 | 6.248 | 1.2553 | -5.1157E-08 | 3.2685E-02 | 3.7597E-02 |
| 57 | 138.880 | 4.855 | 6.268 | 1.3278 | -5.3482E-08 | 3.3910E-02 | 3.9231E-02 |
| 58 | 139.870 | 4.877 | 6.296 | 1.4034 | -5.5906E-08 | 3.5160E-02 | 4.0919E-02 |
| 59 | 140.880 | 4.893 | 6.317 | 1.4820 | -5.8382E-08 | 3.6435E-02 | 4.2661E-02 |
| 60 | 141.870 | 4.915 | 6.345 | 1.5638 | -6.0970E-08 | 3.7735E-02 | 4.4460E-02 |
| 61 | 145.880 | 4.962 | 6.406 | 1.6489 | -6.3738E-08 | 3.9060E-02 | 4.6317E-02 |
| 62 | 151.890 | 5.031 | 6.495 | 1.7373 | -6.6721E-08 | 4.0410E-02 | 4.8234E-02 |
| 63 | 152.890 | 5.041 | 6.508 | 1.8291 | -6.9539E-08 | 4.1785E-02 | 5.0214E-02 |
| 64 | 157.900 | 5.089 | 6.570 | 1.9245 | -7.2618E-08 | 4.3185E-02 | 5.2260E-02 |
| 65 | 158.890 | 5.099 | 6.583 | 2.0235 | -7.5748E-08 | 4.4611E-02 | 5.4372E-02 |
| 66 | 163.890 | 5.083 | 6.562 | 2.1263 | -7.8782E-08 | 4.6061E-02 | 5.6555E-02 |
| 67 | 164.890 | 5.210 | 6.726 | 2.2329 | -8.2859E-08 | 4.7537E-02 | 5.8809E-02 |
| 68 | 165.890 | 5.123 | 6.614 | 2.3436 | -8.6659E-08 | 4.9040E-02 | 6.1140E-02 |
| 69 | 168.900 | 5.192 | 6.703 | 2.4582 | -8.9561E-08 | 5.0566E-02 | 6.3548E-02 |
| 70 | 173.900 | 5.237 | 6.761 | 2.5771 | -9.3656E-08 | 5.2118E-02 | 6.6038E-02 |
| 71 | 174.900 | 5.246 | 6.773 | 2.7003 | -9.7677E-08 | 5.3695E-02 | 6.8612E-02 |
| 72 | 179.910 | 5.290 | 6.829 | 2.8279 | -1.0208E-07 | 5.5297E-02 | 7.1274E-02 |
| 73 | 180.910 | 5.299 | 6.841 | 2.9602 | -1.0648E-07 | 5.6924E-02 | 7.4029E-02 |
| 74 | 183.910 | 5.359 | 6.918 | 3.0971 | -1.1176E-07 | 5.8576E-02 | 7.6878E-02 |
| 75 | 186.910 | 5.351 | 6.908 | 3.2389 | -1.1642E-07 | 6.0252E-02 | 7.9828E-02 |
| 76 | 189.920 | 5.376 | 6.940 | 3.3857 | -1.2177E-07 | 6.1952E-02 | 8.2882E-02 |
| 77 | 192.920 | 5.401 | 6.973 | 3.5377 | -1.2756E-07 | 6.3677E-02 | 8.6045E-02 |
| 78 | 194.920 | 5.418 | 6.995 | 3.6949 | -1.3351E-07 | 6.5425E-02 | 8.9322E-02 |
| 79 | 196.920 | 5.437 | 7.019 | 3.8577 | -1.3992E-07 | 6.7196E-02 | 9.2717E-02 |
| 80 | 201.930 | 5.475 | 7.068 | 4.0260 | -1.4716E-07 | 6.8989E-02 | 9.6237E-02 |
| 81 | 204.930 | 5.483 | 7.079 | 4.2003 | -1.5434E-07 | 7.0805E-02 | 9.9888E-02 |
| 82 | 207.930 | 5.505 | 7.107 | 4.3805 | -1.6234E-07 | 7.2642E-02 | 1.0368E-01 |
| 83 | 208.930 | 5.531 | 7.140 | 4.5669 | -1.7111E-07 | 7.4499E-02 | 1.0761E-01 |
| 84 | 210.00 | 5.539 | 7.151 | 4.7597 | -1.7996E-07 | 7.6377E-02 | 1.1169E-01 |
| 85 | 215.000 | 5.578 | 7.201 | 4.9591 | -1.9067E-07 | 7.8272E-02 | 1.1593E-01 |
| 86 | 222.000 | 5.632 | 7.271 | 5.1652 | -2.0283E-07 | 8.0185E-02 | 1.2033E-01 |
| 87 | 223.000 | 5.640 | 7.281 | 5.3785 | -2.1438E-07 | 8.2115E-02 | 1.2491E-01 |
| 88 | 226.000 | 5.663 | 7.311 | 5.5991 | -2.2763E-07 | 8.4059E-02 | 1.2968E-01 |
| 89 | 227.000 | 5.670 | 7.320 | 5.8272 | -2.4140E-07 | 8.6017E-02 | 1.3464E-01 |
| 90 | 231.000 | 5.707 | 7.368 | 6.0631 | -2.5795E-07 | 8.7985E-02 | 1.3981E-01 |
| 91 | 231.000 | 5.700 | 7.359 | 6.3072 | -2.7374E-07 | 8.9963E-02 | 1.4519E-01 |
| 92 | 238.000 | 5.751 | 7.425 | 6.5594 | -2.9461E-07 | 9.1947E-02 | 1.5081E-01 |
| 93 | 237.000 | 5.744 | 7.415 | 6.8205 | -3.1393E-07 | 9.3936E-02 | 1.5667E-01 |
| 94 | 244.000 | 5.794 | 7.480 | 7.0904 | -3.3950E-07 | 9.5925E-02 | 1.6278E-01 |
| 95 | 243.000 | 5.787 | 7.471 | 7.3698 | -3.6324E-07 | 9.7913E-02 | 1.6917E-01 |
| 96 | 247.000 | 5.816 | 7.508 | 7.6588 | -3.9284E-07 | 9.9894E-02 | 1.7585E-01 |
| 97 | 247.000 | 5.816 | 7.508 | 7.9580 | -4.2298E-07 | 1.0186E-01 | 1.8284E-01 |
| 98 | 251.000 | 5.844 | 7.545 | 8.2675 | -4.5970E-07 | 1.0382E-01 | 1.9017E-01 |
| 99 | 254.000 | 5.865 | 7.572 | 8.5878 | -5.0047E-07 | 1.0575E-01 | 1.9784E-01 |
| 100 | 257.000 | 5.886 | 7.599 | 8.9196 | -5.4564E-07 | 1.0766E-01 | 2.0589E-01 |
| 101 | 258.000 | 5.893 | 7.608 | 9.2632 | -5.9456E-07 | 1.0953E-01 | 2.1435E-01 |
| 102 | 257.000 | 5.886 | 7.599 | 9.6193 | -6.4691E-07 | 1.1136E-01 | 2.2323E-01 |
| 103 | 260.000 | 5.906 | 7.625 | 9.9880 | -7.1139E-07 | 1.1314E-01 | 2.3258E-01 |
| 104 | 255.000 | 5.872 | 7.581 | 10.3707 | -7.7249E-07 | 1.1486E-01 | 2.4244E-01 |
| 105 | 262.000 | 5.920 | 7.643 | 10.7666 | -8.6036E-07 | 1.1650E-01 | 2.5282E-01 |
| 106 | 263.000 | 5.927 | 7.652 | 11.1776 | -9.4908E-07 | 1.1806E-01 | 2.6379E-01 |
| 107 | 262.000 | 5.920 | 7.643 | 11.6044 | -1.0473E-06 | 1.1952E-01 | 2.7539E-01 |
| 108 | 264.000 | 5.934 | 7.661 | 12.0470 | -1.1650E-06 | 1.2085E-01 | 2.8767E-01 |
| 109 | 266.000 | 5.947 | 7.678 | 12.5066 | -1.2985E-06 | 1.2205E-01 | 3.0069E-01 |

| 110 | 268.000 | 5.961 | 7.696 | 12.9841 | -1.4524E-06 | 1.2308E-01 | 3.1451E-01 |
|---|---|---|---|---|---|---|---|

## III. Fine structure of hydrogen-like atoms

### A. Fundamental principle

Under the magnetic potential correction of Eq. (5), the steady-state Schrödinger-like equation of our hydrogen-like atom is

$$\left(-\frac{1}{2\mu}\nabla^2-\frac{Z+\delta_0(Z)}{r}-\frac{\delta_1(Z)}{r^2\sin^2(\theta)}-\frac{\delta_2(Z)}{r^2\cos^2(\theta)}-\frac{\delta_3(Z)}{r^2}\right)\Psi = E\Psi \quad (8).$$

and its solution is

$$\begin{cases} E = E_{\mathrm{M}}(n,l,m,I,J) = -\dfrac{\mu\big(Z+\delta_0(Z)\big)^2}{2\Big(n-l+J-0.5+(-1)^J\sqrt{(L+0.5)^2-2\mu\delta_3(Z)}\Big)^2} \\ \Psi = \varphi_{n,l,m,I,J;\xi}(\vec{\boldsymbol{r}}) = \Phi_m(\phi)\Theta_{l,m,I}(\theta)R_{n,l,m,I,J;\xi}(r) \end{cases} \quad (9).$$

where, the reduced mass of electrons is defined to be $\mu$, the principal quantum number is defined to be $n \in \mathbb{Z}^+$, the angular quantum number is defined to be $l \in \mathbb{N}$ and $l < n$, the magnetic quantum number is defined to be $m \in \mathbb{Z}$ and $|m| \le l$, the newly introduced quantum number is defined to be

$$I = \begin{cases} 0 \text{ or } 1 & \text{If}(m=0) \\ 1 & \text{Else} \end{cases} \quad \text{and} \quad J = \begin{cases} 0 \text{ or } 1 & \text{If}(l=m=I=0) \\ 0 & \text{Else} \end{cases} \quad (10.1).$$

the modified value of angular quantum number is defined to be

$$L \equiv L_{l,m,I} = l-|m|+0.5-\sqrt{0.25-2\mu\delta_2(Z)}-(-1)^I\sqrt{m^2-2\mu\delta_1(Z)} \quad (10.2).$$

the hydrogen-like orbital function is defined to be (non-normalized)

$$\begin{cases} \Phi_m(\phi) = \begin{cases} \cos(m\phi) & \text{If}(m \ge 0) \\ \sin(|m|\phi) & \text{Else} \end{cases} \\ \Theta_{l,m,I}(\theta) = \sin^{(-1)^{I+1}\sqrt{m^2-2\mu\delta_1(Z)}}(\theta) \displaystyle\sum_{i=0}^{\left[\frac{l-|m|}{2}\right]} a_i \cos^{K_{l,m}-2i}(\theta) \\ R_{n,l,m,I,J;\xi}(r) = \displaystyle\sum_{i=0}^{n-l+J-1} b_i r^{i-0.5+(-1)^J\sqrt{(L+0.5)^2-2\mu\delta_3(Z)}} e^{-\xi r} \end{cases} \quad (10.3).$$

and the coefficient is defined to be

$$\begin{cases} \xi = \dfrac{\mu\big(Z+\delta_0(Z)\big)}{n-l+J-0.5+(-1)^J\sqrt{(L+0.5)^2-2\mu\delta_3(Z)}} \\ K_{l,m} = l-|m|+0.5-\sqrt{0.25-2\mu\delta_2(Z)} \\ a_{i+1} = -\dfrac{\big(K_{l,m}-2i\big)\big(K_{l,m}-2i-1\big)+2\mu\delta_2(Z)}{2(i+1)(2L-2i-1)}a_i \qquad (a_0=b_0=1) \\ b_{i+1} = \dfrac{\Big(2i+1+(-1)^J\sqrt{(2L+1)^2-8\mu\delta_3(Z)}\Big)\xi-2\mu\big(Z+\delta_0(Z)\big)}{(i+1)\Big(i+1+(-1)^J\sqrt{(2L+1)^2-8\mu\delta_3(Z)}\Big)}b_i \end{cases} \quad (10.4).$$

## B. Numerical experiments and conclusions

In Fig.2, our maximum error rate comes from $3D_{5/2}-3D_{3/2}$, which is 13.4%, but other energy level intervals can be consistent with the fine structure in quantum electrodynamics[12], which is superior to Sommerfeld model[10] (the energy level formula of single electron Dirac equation is the same as Sommerfeld model), as shown in Table II and Fig.3. Therefore, in the ab initio calculation of the energy level fine structure of atoms and molecules, the two-step method in relativistic quantum mechanics can be replaced by the solution of a Schrödinger-like equation—In modern computer programs of quantum mechanics and quantum chemistry, the approximate solution of the stationary Schrodinger equation under Coulomb action potential is calculated first, and then its energy level is corrected by its wave function, which is called the two-step method, as shown in the refs[8,9,28] , omitted.

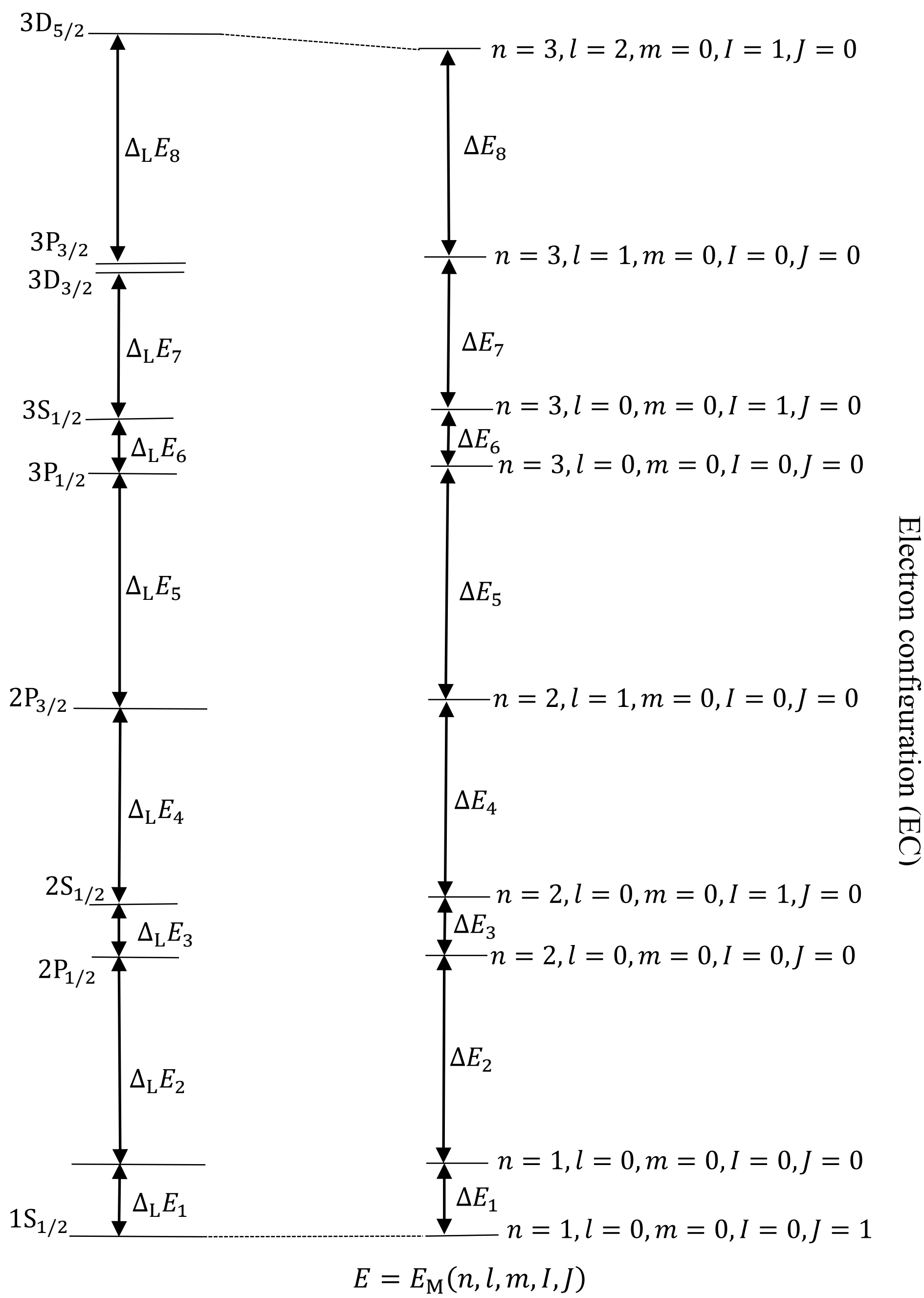

Fig.2. The schematic diagram of energy level interval of hydrogen-like atoms ($1 \leq Z \leq 110$), but excluding hyperfine structures and our redundant energy levels (Not to scale).

Table II. Energy level interval of hydrogen atom (MHz) in the both approximations, and their ionization energies are all 3288097385.73 MHz.

| Lamb shift based on Sommerfeld model | | Ours | | Error rate (%) |
|---|---|---|---|---|
| $\Delta_L E_1$ | 8173 | $\Delta E_1$ | 8166.48 | 0.08 |

| $\Delta_{\mathrm{L}}E_2$ | 2466061413 | $\Delta E_2$ | 2466065094 | 0.00 |
|---|---|---|---|---|
| $\Delta_{\mathrm{L}}E_3$ | 1056.85 | $\Delta E_3$ | 1057.01 | 0.02 |
| $\Delta_{\mathrm{L}}E_4$ | 9911.2 | $\Delta E_4$ | 9880.73 | 0.31 |
| $\Delta_{\mathrm{L}}E_5$ | 456674888.9 | $\Delta E_5$ | 456671701.2 | 0.00 |
| $\Delta_{\mathrm{L}}E_6$ | 314.82 | $\Delta E_6$ | 313.19 | 0.52 |
| $\Delta_{\mathrm{L}}E_7$ | 2929.9 | $\Delta E_7$ | 2927.63 | 0.08 |
| $\Delta_{\mathrm{L}}E_8$ | 1110 | $\Delta E_8$ | 961.33 | 13.4 |

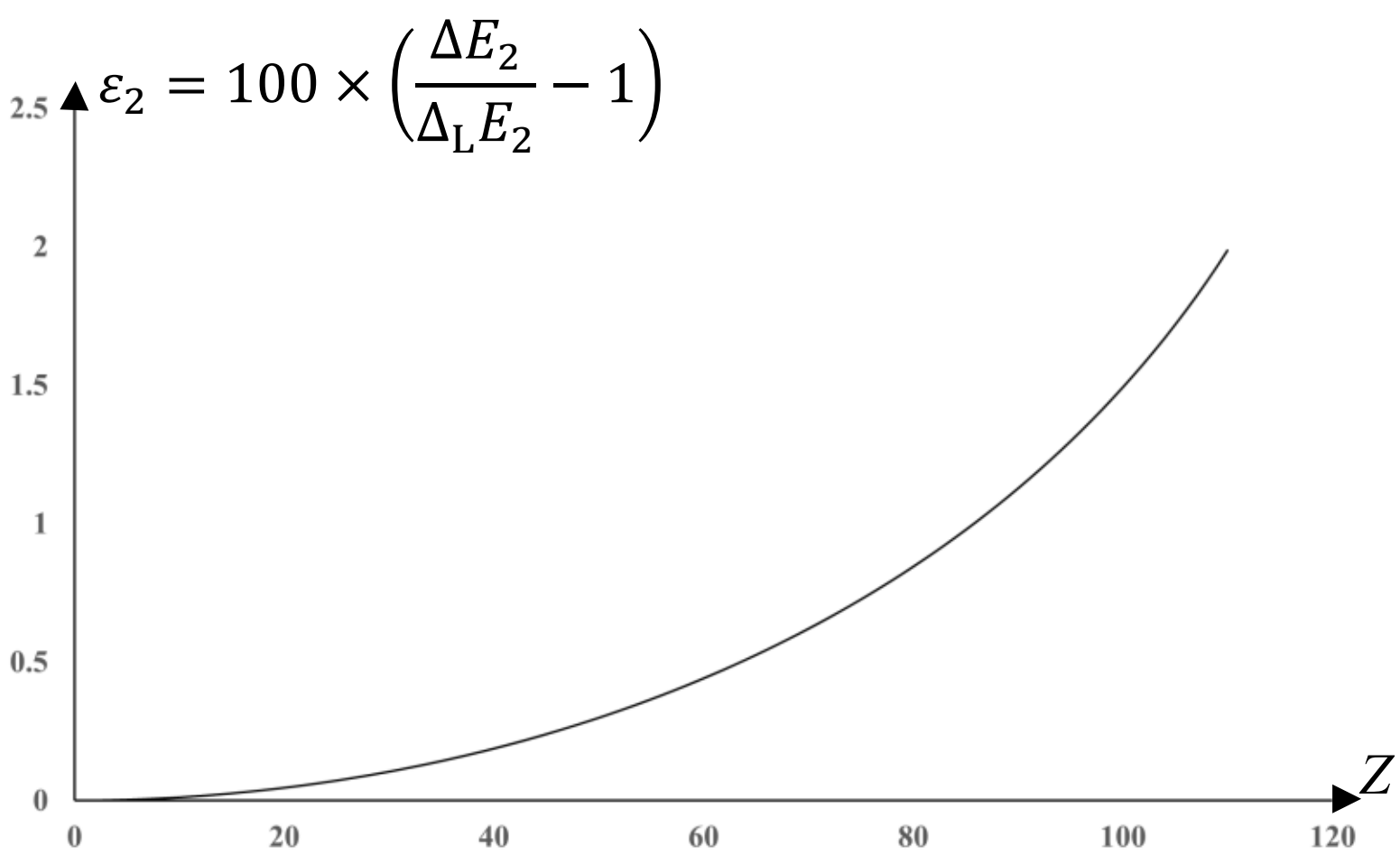

Fig.3. Error rate of energy level interval two

## IV. Asymmetric variational method (HAV) based on hydrogen-like atom orbit

### A. Fundamental principle

Let the serial number of electron configuration be denoted as

$$[\![k]\!] = \left(n_{k,1}, l_{k,1}, m_{k,1}, I_{k,1}, \xi_{k,1}; n_{k,2}, l_{k,2}, m_{k,2}, I_{k,2}, \xi_{k,2} \cdots\right) \quad (11.1).$$

the asymmetric variational function of Hartree type is written as

$${}^{s}\psi_{[\![k]\!]} = \prod_{i=1}^{N} \Phi_{m_{k,i}}(\phi_i)\Theta_{l_{k,i},m_{k,i},I_{k,i}}(\theta_i) R_{n_{k,i},l_{k,i},m_{k,i},I_{k,i},0;\ {}^{s}\xi_{k,i}}(r_i) \quad (11.2).$$

the basis set function of even terms is recorded as

$$
{}^{s}\omega_{[\![k]\!]} = 1 + \sum_{i=1}^{N-1}\sum_{j=i+1}^{N}\left( {}^{s}\lambda_{k,1}\,{}^{s}A_{k,i,j}r_{i,j}^{2} + {}^{s}\lambda_{k,2}\,{}^{s}B_{k,i,j}\left(r_i - r_j\right)^2 + {}^{s}\lambda_{k,3}\,{}^{s}C_{k,i,j}r_ir_j\right) \quad (11.3).
$$

the normalized coefficients in the basis set are recorded as

$$
\begin{cases}
{}^{s}A_{k,i,j} = \dfrac{1}{\sqrt{\left\langle {}^{s}\psi_{[\![k]\!]} \middle| r_{i,j}^{4} \middle| {}^{s}\psi_{[\![k]\!]} \right\rangle}}\sqrt{\left\langle {}^{s}\psi_{[\![k]\!]} \middle| {}^{s}\psi_{[\![k]\!]} \right\rangle} \\
{}^{s}B_{k,i,j} = \dfrac{1}{\sqrt{\left\langle {}^{s}\psi_{[\![k]\!]} \middle| \left(r_i - r_j\right)^{4} \middle| {}^{s}\psi_{[\![k]\!]} \right\rangle}}\sqrt{\left\langle {}^{s}\psi_{[\![k]\!]} \middle| {}^{s}\psi_{[\![k]\!]} \right\rangle} \\
{}^{s}C_{k,i,j} = \dfrac{1}{\sqrt{\left\langle {}^{s}\psi_{[\![k]\!]} \middle| r_{i}^{2}r_{j}^{2} \middle| {}^{s}\psi_{[\![k]\!]} \right\rangle}}\sqrt{\left\langle {}^{s}\psi_{[\![k]\!]} \middle| {}^{s}\psi_{[\![k]\!]} \right\rangle}
\end{cases} \quad (11.4).
$$

our variational function be denoted as

$$
{}^{s}\Psi_{[\![k]\!]} = {}^{s}\psi_{[\![k]\!]}\,{}^{s}\omega_{[\![k]\!]}e^{-\mathrm{i}Et} \quad (11.5).
$$

then the objective function of the atomic reference state in our HAV method is

$$
\begin{aligned}
E = \mathrm{Min}\ {}^{0}E\left({}^{0}\vec{\boldsymbol{\xi}}_k, {}^{0}\vec{\boldsymbol{\lambda}}_k\right) &= \frac{1}{\left\langle {}^{0}\Psi_{[\![k]\!]} \middle| {}^{0}\Psi_{[\![k]\!]} \right\rangle}\left\langle {}^{0}\Psi_{[\![k]\!]} \middle| \hat{H} \middle| {}^{0}\Psi_{[\![k]\!]} \right\rangle \\
s.t.\quad n_{k,i} = l_{k,i} + 1 \ \text{ and } \ m_{k,i} &= 0 \quad (i = 1,2\cdots)
\end{aligned} \quad (12.1).
$$

the objective function of atomic triplet is

$$
\begin{aligned}
&E = \mathrm{Min}\ {}^{3}E\left({}^{3}\vec{\boldsymbol{\xi}}_k\right) = \frac{1}{\left\langle {}^{3}\Psi_{[\![k]\!]} \middle| {}^{3}\Psi_{[\![k]\!]} \right\rangle}\left\langle {}^{3}\Psi_{[\![k]\!]} \middle| \hat{H} \middle| {}^{3}\Psi_{[\![k]\!]} \right\rangle \\
&s.t.\quad \begin{cases}
l_{k,i} = l_{p,i} \ \text{ and } \ m_{k,i} = m_{p,i} \quad (i = 1,2\cdots) \\
{}^{3}\lambda_{k,1} = -\dfrac{\left\langle {}^{3}\psi_{[\![k]\!]} \middle| {}^{0}\Psi_{[\![p]\!]} \right\rangle}{\left\langle {}^{3}\psi_{[\![k]\!]} \middle| \sum_{i\geq 1}\sum_{j>i}\ {}^{3}A_{k,i,j}r_{i,j}^{2} \middle| {}^{0}\Psi_{[\![p]\!]} \right\rangle} \\
{}^{3}\lambda_{k,2} = {}^{3}\lambda_{k,3} = 0
\end{cases}
\end{aligned} \quad (12.2).
$$

the objective function of atomic singlet is

$$
\begin{aligned}
&E = \mathrm{Min}\ {}^{1}E\left({}^{1}\vec{\boldsymbol{\xi}}_k\right) = \frac{1}{\left\langle {}^{1}\Psi_{[\![k]\!]} \middle| {}^{1}\Psi_{[\![k]\!]} \right\rangle}\left\langle {}^{1}\Psi_{[\![k]\!]} \middle| \hat{H} \middle| {}^{1}\Psi_{[\![k]\!]} \right\rangle \\
&s.t.\quad \begin{cases}
n_{k,i} = n_{q,i}, l_{k,i} = l_{p,i} = l_{q,i}, \text{and } m_{k,i} = m_{p,i} = m_{q,i} \quad (i = 1,2\cdots) \\
{}^{1}\lambda_{k,1} = \dfrac{f_{p,0}f_{q,2} - f_{q,0}f_{p,2}}{f_{q,1}f_{p,2} - f_{p,1}f_{q,2}} \\
{}^{1}\lambda_{k,2} = -\dfrac{1}{f_{p,2}}\left(f_{p,0} + {}^{1}\lambda_{k,1}f_{p,1}\right) \\
{}^{1}\lambda_{k,3} = 0
\end{cases}
\end{aligned} \quad (12.3).
$$

where, the $[\![p]\!]$ is reference state corresponding to $[\![k]\!]$, the $[\![q]\!]$ is triplet state corresponding to $[\![k]\!]$, and the substitution function in singlet constraint equation is defined to be

$$f_{p,0} = f_{p,0}\left( {}^{1}\vec{\xi}_k \right) = \left\langle {}^{1}\psi_{[\![k]\!]} \middle| {}^{0}\Psi_{[\![p]\!]} \right\rangle \quad (13.1).$$

$$f_{p,1} = f_{p,1}\left( {}^{1}\vec{\xi}_k \right) = \left\langle {}^{1}\psi_{[\![k]\!]} \middle| \sum_{i\geq 1}\sum_{j>i} {}^{1}A_{k,i,j} r_{i,j}^{2} \middle| {}^{0}\Psi_{[\![p]\!]} \right\rangle \quad (13.2).$$

$$f_{p,2} = f_{p,2}\left( {}^{1}\vec{\xi}_k \right) = \left\langle {}^{1}\psi_{[\![k]\!]} \middle| \sum_{i\geq 1}\sum_{j>i} {}^{1}B_{k,i,j} \left(r_i - r_j\right)^{2} \middle| {}^{0}\Psi_{[\![p]\!]} \right\rangle \quad (13.3).$$

$$f_{q,0} = f_{q,0}\left( {}^{1}\vec{\xi}_k \right) = \left\langle {}^{1}\psi_{[\![k]\!]} \middle| {}^{3}\Psi_{[\![q]\!]} \right\rangle \quad (13.4).$$

$$f_{q,1} = f_{q,1}\left( {}^{1}\vec{\xi}_k \right) = \left\langle {}^{1}\psi_{[\![k]\!]} \middle| \sum_{i\geq 1}\sum_{j>i} {}^{1}A_{k,i,j} r_{i,j}^{2} \middle| {}^{3}\Psi_{[\![q]\!]} \right\rangle \quad (13.5).$$

$$f_{q,2} = f_{q,2}\left( {}^{1}\vec{\xi}_k \right) = \left\langle {}^{1}\psi_{[\![k]\!]} \middle| \sum_{i\geq 1}\sum_{j>i} {}^{1}B_{k,i,j} \left(r_i - r_j\right)^{2} \middle| {}^{3}\Psi_{[\![q]\!]} \right\rangle \quad (13.6).$$

In addition, the reference state is a possible mathematical approximate solution in Schrödinger equation, which is introduced to construct the constraint equation in the objective function. Therefore, its electronic configuration may not satisfy the Pauli exclusion principle, so it is a possible redundant solution, that is, there is no experimental observation corresponding to it.

### B. (Non-optimized) Some useful integrals algorithms

#### *1. Single electron integral*

Input: $\left(\vec{\boldsymbol{k}}, \vec{\xi}, n_1, l_1, m_1, I_1, n_2, l_2, m_2, I_2, i\right)$;

Output: $X = \int_0^{+\infty}\int_0^{\pi}\int_0^{2\pi} Y^{*(1)}\left(\vec{\boldsymbol{k}}; \vec{\boldsymbol{\theta}}, \vec{\boldsymbol{\phi}}\right)\sin(\theta_1) r_1^{k_5+2} \varphi_{[\![1]\!]}^{*(0)}(\vec{\boldsymbol{r}}_1)\varphi_{[\![2]\!]}^{*(i)}(\vec{\boldsymbol{r}}_1)\, dr d\theta_1 d\phi_1$;

Algorithmic process:

$$F_e\left(\vec{\boldsymbol{k}}, \vec{\xi}, n_1, l_1, m_1, I_1, n_2, l_2, m_2, I_2, i\right)\{$$

$$\text{If}(m_1 \geq 0)\{p_1 \leftarrow 0;\}\ \text{Else}\{p_1 \leftarrow 1;\}$$

$$\text{If}(m_2 \geq 0)\{p_2 \leftarrow 0;\}\ \text{Else}\{p_2 \leftarrow 1;\}$$

$$x_1 \leftarrow 0; \text{For}\left(s_1 \leftarrow 0; s_1 \leq \left[\frac{|m_1| - p_1}{2}\right]; s_1 \leftarrow s_1 + 1\right)\{$$

$$\mathrm{For}\left(s_2 \leftarrow 0; s_2 \leq \left[\frac{|m_2| - p_2}{2}\right]; s_2 \leftarrow s_2 + 1\right)\{$$

$$y \leftarrow \frac{(-1)^{s_1+s_2} 2^{|m_1|+|m_2|-2s_1-2s_2-2} (|m_1| - s_1 - 1)!\,(|m_2| - s_2 - 1)!}{(|m_1| - 2s_1 - p_1)!\,(|m_2| - 2s_2 - p_2)!\,s_1!\,s_2!};$$

$$x_1 \leftarrow x_1 + y \int_0^{2\pi} \sin^{p_1+p_2+k_3}(\phi) \cos^{|m_1|+|m_2|-2s_1-2s_2-p_1-p_2+k_4}(\phi) d\phi\,;$$

$$\}\}\ \ \mathrm{If}(m_1 > 0)\{x_1 \leftarrow m_1 x_1;\}\ \mathrm{If}(m_2 > 0)\{x_1 \leftarrow m_2 x_1;\}$$

$$K_1 \leftarrow l_1 - |m_1| + 0.5 - \sqrt{0.25 - 2\delta_2(Z)};$$

$$K_2 \leftarrow l_2 - |m_2| + 0.5 - \sqrt{0.25 - 2\delta_2(Z)};$$

$$L_1 \leftarrow K_1 - (-1)^{I_1} \sqrt{m_1^2 - 2\delta_1(Z)};$$

$$L_2 \leftarrow K_2 - (-1)^{I_2} \sqrt{m_2^2 - 2\delta_1(Z)};$$

$$x_2 \leftarrow 0; \mathrm{For}\left(s_1 \leftarrow 0; s_1 \leq \left[\frac{l_1 - |m_1|}{2}\right]; s_1 \leftarrow s_1 + 1\right)\{$$

$$\mathrm{For}\left(s_2 \leftarrow 0; s_2 \leq \left[\frac{l_2 - |m_2|}{2}\right]; s_2 \leftarrow s_2 + 1\right)\{$$

$$\mathrm{If}(s_1 = 0)\{a_{1,0} \leftarrow 1;\}$$

$$\mathrm{Else}\left\{a_{1,s_1} \leftarrow -\frac{(K_1 - 2s_1 + 2)(K_1 - 2s_1 + 1) + 2\delta_2(Z)}{2s_1(2L_1 - 2s_1 + 1)} a_{1,s_1-1};\right\}$$

$$\mathrm{If}(s_2 = 0)\{a_{2,0} \leftarrow 1;\}$$

$$\mathrm{Else}\left\{a_{2,s_2} \leftarrow -\frac{(K_2 - 2s_2 + 2)(K_2 - 2s_2 + 1) + 2\delta_2(Z)}{2s_2(2L_2 - 2s_2 + 1)} a_{2,s_2-1};\right\}$$

$$y \leftarrow a_{1,s_1} a_{2,s_2} \int_0^{\pi} \sin^{|m_1|+|m_2|+k_1+1}(\theta) \cos^{l_1+l_2-|m_1|-|m_2|+k_2-2s_1-2s_2}(\theta) d\theta\,;$$

$$\mathrm{If}(i \neq 1)\{x_2 \leftarrow x_2 + y;\}\ \mathrm{Else}\ \{x_2 \leftarrow x_2 + (K_2 - 2s_2)y;\}$$

$$\}\}\ \ x_3 \leftarrow 0;\ \mathrm{For}(s_1 \leftarrow 0; s_1 \leq n_1 - l_1 - 1; s_1 \leftarrow s_1 + 1)\{$$

$$\mathrm{For}(s_2 \leftarrow 0; s_2 \leq n_2 - l_2 - 1; s_2 \leftarrow s_2 + 1)\{$$

$$\mathrm{If}(s_1 = 0)\{b_{1,0} \leftarrow 1;\}$$

$$\text{Else}\left\{b_{1,s_1} \leftarrow \frac{\left(2s_1-1+\sqrt{(2L_1+1)^2-8\delta_3(Z)}\right)\xi_1-2Z-2\delta_0(Z)}{s_1\left(s_1+\sqrt{(2L_1+1)^2-8\delta_3(Z)}\right)} b_{1,s_1-1};\right\}$$

$$\text{If}(s_2 = 0)\{b_{2,0} \leftarrow 1;\}$$

$$\text{Else}\left\{b_{2,s_2} \leftarrow \frac{\left(2s_2-1+\sqrt{(2L_2+1)^2-8\delta_3(Z)}\right)\xi_2-2Z-2\delta_0(Z)}{s_2\left(s_2+\sqrt{(2L_2+1)^2-8\delta_3(Z)}\right)} b_{2,s_2-1};\right\}$$

$$\text{If}(i \le 1)\{y \leftarrow 1;\}\ \text{Else}\left\{y \leftarrow s_2 - 0.5 + \sqrt{(L_2+0.5)^2 - 2\delta_3(Z)};\right\}$$

$$x_3 \leftarrow x_3 + b_{1,s_1} b_{2,s_2} y \frac{(l_1+l_2+s_1+s_2+k_5+2)!}{(\xi_1+\xi_2)^{l_1+l_2+s_1+s_2+k_5+3}};$$

$$\}\}\quad \text{Return}\ X \leftarrow x_1x_2x_3;\} \qquad\qquad // \text{End.}$$

where,
$$Y^{*(s)}\left(\vec{\boldsymbol{k}};\vec{\boldsymbol{\theta}},\vec{\boldsymbol{\phi}}\right) = \prod_{i=1}^{s} \sin^{k_{4i-3}}(\theta_i)\cos^{k_{4i-2}}(\theta_i)\sin^{k_{4i-1}}(\phi_i)\cos^{k_{4i}}(\phi_i) \quad (14.1).$$
$$s.t.\quad k_{4i-3} \ge k_{4i-1} + k_{4i}\ \text{ and }\ k_1 \ge \text{Max}\{k_1, k_5, \cdots\}$$

$$\varphi^{*(i)}_{[\![k]\!]}(\vec{\boldsymbol{r}}) = \begin{cases} \Phi_{m_k}(\phi)\Theta_{l_k,m_k,I_k}(\theta)R_{n_k,l_k,m_k,I_k,0;\xi_k}(r) & \text{If}(i=0) \\ \Phi_{m_k}(\phi)\Theta^{*}_{l_k,m_k,I_k}(\theta)R_{n_k,l_k,m_k,I_k,0;\xi_k}(r) & \text{Else If}(i=1) \\ \Phi_{m_k}(\phi)\Theta_{l_k,m_k,I_k}(\theta)R^{*}_{n_k,l_k,m_k,I_k,0;\xi_k}(r) & \text{Else} \end{cases} \quad (14.2).$$

$$\Theta^{*}_{l,m,I}(\theta) = \sin^{(-1)^{I+1}\sqrt{m^2-2\delta_1(Z)}}(\theta) \sum_{i=0}^{\left[\frac{l-|m|}{2}\right]} \left(K_{l,m} - 2i\right)a_i\cos^{K_{l,m}-2i}(\theta) \quad (14.3).$$

$$R^{*}_{n,l,m,I,0;\xi}(r) = \sum_{i=0}^{n-l-1}\left(i - 0.5 + \sqrt{(L+0.5)^2 - 2\delta_3(Z)}\right)b_i r^{i-0.5+\sqrt{(L+0.5)^2-2\delta_3(Z)}}e^{-\xi r} \quad (14.4).$$

## ***2. Volume element***

$$\text{Let}\begin{cases} x_1 = r_1\sin(\theta)\cos(\phi) \\ y_1 = r_1\sin(\theta)\sin(\phi) \\ z_1 = r_1\cos(\theta) \\ \dfrac{z_2}{r_2} = \cos(\theta)\cos(\beta) + \sin(\theta)\sin(\beta)\cos(\chi) \\ \dfrac{x_2}{r_2} = \dfrac{\cos(\beta)\cos(\phi)}{\sin(\theta)} + \sin(\beta)\sin(\chi)\sin(\phi) - \dfrac{z_2}{r_2}\dfrac{\cos(\theta)\cos(\phi)}{\sin(\theta)} \\ \dfrac{y_2}{r_2} = \dfrac{\cos(\beta)sin(\phi)}{\sin(\theta)} - \sin(\beta)\sin(\chi)\cos(\phi) - \dfrac{z_2}{r_2}\dfrac{\cos(\theta)\sin(\phi)}{\sin(\theta)} \end{cases} \quad (15.1).$$

then you can get:

$$dx_1dy_1dz_1\,dx_2dy_2dz_2 = \begin{vmatrix} \frac{\partial}{\partial r_1}z_2 & \frac{\partial}{\partial r_2}z_2 & \frac{\partial}{\partial r_{1,2}}z_2 & \frac{\partial}{\partial \theta}z_2 & \frac{\partial}{\partial \phi}z_2 & \frac{\partial}{\partial \chi}z_2 \\ \frac{\partial}{\partial r_1}y_2 & \frac{\partial}{\partial r_2}y_2 & \frac{\partial}{\partial r_{1,2}}y_2 & \frac{\partial}{\partial \theta}y_2 & \frac{\partial}{\partial \phi}y_2 & \frac{\partial}{\partial \chi}y_2 \\ \frac{\partial}{\partial r_1}x_2 & \frac{\partial}{\partial r_2}x_2 & \frac{\partial}{\partial r_{1,2}}x_2 & \frac{\partial}{\partial \theta}x_2 & \frac{\partial}{\partial \phi}x_2 & \frac{\partial}{\partial \chi}x_2 \\ \frac{\partial}{\partial r_1}z_1 & \frac{\partial}{\partial r_2}z_1 & \frac{\partial}{\partial r_{1,2}}z_1 & \frac{\partial}{\partial \theta}z_1 & \frac{\partial}{\partial \phi}z_1 & \frac{\partial}{\partial \chi}z_1 \\ \frac{\partial}{\partial r_1}y_1 & \frac{\partial}{\partial r_2}y_1 & \frac{\partial}{\partial r_{1,2}}y_1 & \frac{\partial}{\partial \theta}y_1 & \frac{\partial}{\partial \phi}y_1 & \frac{\partial}{\partial \chi}y_1 \\ \frac{\partial}{\partial r_1}x_1 & \frac{\partial}{\partial r_2}x_1 & \frac{\partial}{\partial r_{1,2}}x_1 & \frac{\partial}{\partial \theta}x_1 & \frac{\partial}{\partial \phi}x_1 & \frac{\partial}{\partial \chi}x_1 \end{vmatrix}$$

$$= r_1r_2r_{1,2}\sin(\theta)dr_1dr_2dr_{1,2}d\theta d\phi d\chi \qquad \left(\beta = \mathrm{Arccos}\left(\frac{r_1^2 + r_2^2 - r_{1,2}^2}{2r_1r_2}\right)\right) \quad (15.2).$$

### *3. Angular integration of single-center two electrons*

Input: $(\vec{\boldsymbol{k}})$;

Output: $(\vec{\boldsymbol{a}})$;

$$s.t. \sum_{i\geq 0}\sum_{j\geq 0} a_{i,j}\sin^i(\beta)\cos^j(\beta) = \int_0^{\pi} d\theta \int_0^{2\pi} d\phi \int_0^{2\pi} \sin(\theta) f_{\mathrm{A}}(\vec{\boldsymbol{k}}; \beta, \theta, \phi, \chi) d\chi\,;$$

Algorithmic process:

$F_{\mathrm{A}}(\vec{\boldsymbol{k}})\{$

$\mathrm{For}(i \leftarrow 0; i \leq k_5 + k_6 + v; i \leftarrow i + 1)\{$

$\mathrm{For}(j \leftarrow 0; j \leq k_5 + k_6 + v; j \leftarrow j + 1)\{$

$a_{i,j} \leftarrow 0;$

$\}\}\quad v \leftarrow (k_5 - k_8 - k_7) \bmod 2;$

$\mathrm{For}(i_1 \leftarrow 0; i_1 \leq k_8; i_1 \leftarrow i_1 + 1)\{$

$\mathrm{For}(i_2 \leftarrow 0; i_2 \leq k_8 - i_1; i_2 \leftarrow i_2 + 1)\{$

$\mathrm{For}(i_3 \leftarrow 0; i_3 \leq k_7; i_3 \leftarrow i_3 + 1)\{$

$\mathrm{For}(i_4 \leftarrow 0; i_4 \leq k_7 - i_3; i_4 \leftarrow i_4 + 1)\{$

$\mathrm{For}(i_5 \leftarrow 0; i_5 \leq v; i_5 \leftarrow i_5 + 1)\{$

$\mathrm{For}\left(i_6 \leftarrow 0; i_6 \leq \frac{k_5 - k_8 - k_7 - v}{2}; i_6 \leftarrow i_6 + 1\right)\{$

$$\text{For}(i_7 \leftarrow 0; i_7 \le k_6 + i_2 + i_4 + 2i_5 + 2i_6; i_7 \leftarrow i_7 + 1)\{$$

$$x \leftarrow \frac{(-1)^{i_2+i_3+i_4+i_6} k_8!\, k_7!\, (2i_5)!}{4^{i_5}(1-2i_5) i_1!\, i_2!\, i_3!\, i_4!\, (i_5!)^2 i_6!\, i_7!\, (k_8 - i_1 - i_2)!};$$

$$x \leftarrow \frac{\left(\frac{k_5-k_8-k_7-v}{2}\right)!\, (k_6 + i_2 + i_4 + 2i_5 + 2i_6)!}{(k_7 - i_3 - i_4)!\left(\frac{k_5-k_8-k_7-v}{2} - i_6\right)!\, (k_6 + i_2 + i_4 + 2i_5 + 2i_6 - i_7)!} x;$$

$$x \leftarrow x \int_0^{\pi} \sin^{k_1-k_8-k_7+i_1+i_3+i_7+1}(\theta) \cos^{k_2+k_6+2i_2+2i_4+2i_5+2i_6-i_7}(\theta) d\theta\,;$$

$$x \leftarrow x \int_0^{2\pi} \sin^{k_3+k_7+i_1-i_3}(\phi) \cos^{k_4+k_8-i_1+i_3}(\phi) d\phi \int_0^{2\pi} \sin^{i_1+i_3+i_7}(\chi) d\chi\,;$$

$$i \leftarrow i_1 + i_3 + i_7;\ \ j \leftarrow k_8 + k_7 + k_6 - i_1 - i_3 + 2i_5 + 2i_6 - i_7;$$

$$a_{i,j} \leftarrow a_{i,j} + x;$$

$$\}\}\}\}\}\}\}\quad \text{Return}\ \ \vec{\boldsymbol{a}};\}$$ //End.

where
$$\underset{v=(k_5-k_8-k_7) \bmod 2}{f_{\mathrm{A}}\left(\vec{\boldsymbol{k}}; \beta, \theta, \phi, \chi\right)} = \sum_{i_1=0}^{k_8} \sum_{i_2=0}^{k_8-i_1} \sum_{i_3=0}^{k_7} \sum_{i_4=0}^{k_7-i_3} \sum_{i_5=0}^{v} \sum_{i_6=0}^{\left\lceil \frac{k_5-k_8-k_7-v}{2} \right\rceil}$$

$$\sum_{i_7=0}^{k_6+i_2+i_4+2i_5+2i_6} \frac{(-1)^{i_2+i_3+i_4+i_6} k_8!\, k_7!\, (2i_5)!}{4^{i_5}(1-2i_5) i_1!\, i_2!\, i_3!\, i_4!\, (i_5!)^2 i_6!\, i_7!\, (k_8 - i_1 - i_2)!}$$

$$\times \frac{\left(\frac{k_5-k_8-k_7-v}{2}\right)!\, (k_6 + i_2 + i_4 + 2i_5 + 2i_6)!}{(k_7 - i_3 - i_4)!\left(\frac{k_5-k_8-k_7-v}{2} - i_6\right)!\, (k_6 + i_2 + i_4 + 2i_5 + 2i_6 - i_7)!}$$

$$\times \sin^{i_1+i_3+i_7}(\beta) \cos^{k_8+k_7+k_6-i_1-i_3+2i_5+2i_6-i_7}(\beta) \sin^{i_1+i_3+i_7}(\chi)$$

$$\times \sin^{k_1-k_8-k_7+i_1+i_3+i_7}(\theta) \cos^{k_2+k_6+2i_2+2i_4+2i_5+2i_6-i_7}(\theta)$$

$$\times \sin^{k_3+k_7+i_1-i_3}(\phi) \cos^{k_4+k_8-i_1+i_3}(\phi) \qquad (16).$$

### ***4. Radial integration of single-center two electrons***

Input: $(\vec{\boldsymbol{k}}, \vec{\boldsymbol{\xi}})$;

Output: $X = \int_0^{+\infty} r_2^{k_2} e^{-\xi_2 r_2}\ dr_2 \int_{r_2}^{+\infty} r_1^{k_1} e^{-\xi_1 r_1}\ dr_1$ ;

Algorithmic process:

$$F_{\mathrm{R(II)}}\left(\vec{\boldsymbol{k}}, \vec{\boldsymbol{\xi}}\right)\{$$

$$\text{If}(k_2 < 0\ \text{ or }\ k_1 + k_2 + 1 < 0)\{\ \text{Return}\ \ '\mathit{Nonconvergence}';\}$$

$$\text{Else}\{\quad X \leftarrow 0\,;\ \ \text{If}(k_1 \ge 0)\{$$

$$\text{For}(i \leftarrow 0; i \leq k_1; i \leftarrow i+1)\{$$

$$X \leftarrow X + \frac{k_1!\,(k_1+k_2-i)!}{(k_1-i)!\,\xi_1^{i+1}(\xi_1+\xi_2)^{k_1+k_2-i+1}}\ ;$$

$$\}\}\ \ \text{Else}\ \{$$

$$X \leftarrow 0\ ;\ \text{For}(i \leftarrow 0; i \leq k_2; i \leftarrow i+1)\{$$

$$x \leftarrow (-1)^i \frac{(k_1+k_2+1)!\,k_2!}{i!\,(k_2-i)!}\frac{\xi_1^i}{\xi_2^{k_2+1}};$$

$$\text{If}(k_1+2+i=0)\{X \leftarrow X+\xi_2 x\ ;\}$$

$$\text{Else If}(k_1+i+1=0)\left\{X \leftarrow X+x\ \cdot \ln\left(1+\frac{\xi_2}{\xi_1}\right);\right\}$$

$$\text{Else}\left\{X \leftarrow X+\frac{x}{k_1+i+1}\left(\frac{1}{\xi_1^{k_1+1+i}}-\frac{1}{(\xi_1+\xi_2)^{k_1+1+i}}\right);\right\}$$

$$\}\}\quad \text{Return}\quad X;\}\}$$ //End.

## *5. Coupling integration of single-center two electrons*

Input: $(\vec{\boldsymbol{k}},\vec{\boldsymbol{\xi}})$;

Output: $X = \left\langle Y^{*(2)}\left(\vec{\boldsymbol{k}};\vec{\boldsymbol{\theta}},\vec{\boldsymbol{\phi}}\right)\middle| r_1^{k_9} r_2^{k_{10}} r_{1,2}^{k_{11}} e^{-\xi_1 r_1-\xi_2 r_2}\right\rangle$;

Algorithmic process:

$$F_{ee}\left(\vec{\boldsymbol{k}},\vec{\boldsymbol{\xi}}\right)\{$$

$$X \leftarrow 0;$$

$$\text{If}(0 \equiv k_{11} \bmod 2)\{$$

$$\text{For}(i_1 \leftarrow 0; i_1 \leq 0.5k_{11}; i_1 \leftarrow i_1+1)\{$$

$$\text{For}(i_2 \leftarrow 0; i_2 \leq i_1; i_2 \leftarrow i_2+1)\{\ \text{For}(i_3 \leftarrow 0; i_3 \leq i_2; i_3 \leftarrow i_3+1)\{$$

$$x \leftarrow \int_0^{2\pi} \sin^{k_7+i_3}(\phi)\cos^{k_8+i_2-i_3}(\phi)d\phi\ ;$$

$$x \leftarrow x\int_0^{2\pi} \sin^{k_3+i_3}(\phi)\cos^{k_4+i_2-i_3}(\phi)d\phi\ ;$$

$$x \leftarrow x\int_0^{\pi} \sin^{k_5+i_2+1}(\theta)\cos^{k_6+i_1-i_2}(\theta)d\theta\ ;$$

$$x \leftarrow x \int_0^{\pi} \sin^{k_1+i_2+1}(\theta) \cos^{k_2+i_1-i_2}(\theta) d\theta \,;$$

$$\text{If}(x \neq 0)\{$$

$$\text{For}(i_4 \leftarrow 0; i_4 \leq 0.5k_{11} - i_1; i_4 \leftarrow i_4 + 1)\{$$

$$y \leftarrow \frac{(-2)^{i_1}(0.5k_{11})!}{i_3!\, i_4!\, (i_1 - i_2)!\, (i_2 - i_3)!\, (0.5k_{11} - i_1 - i_4)!} x;$$

$$X \leftarrow X + \frac{(k_9 + k_{11} - i_1 - 2i_4 + 2)!\, (k_{10} + i_1 + 2i_4 + 2)!}{\xi_1^{k_9+k_{11}-i_1-2i_4+3} \xi_2^{k_{10}+i_1+2i_4+3}} y;$$

$$\}\}\}\}\}\quad \text{Return}\ \ X; \}\ \text{Else}\{$$

$$\text{If}(k_1 < k_5)\big\{\text{For}(i \leftarrow 1; i \leq 4; i \leftarrow i + 1)\{s \leftarrow k_i; k_i \leftarrow k_{i+4}; k_{i+4} \leftarrow s;\}\big\}$$

$$\vec{\boldsymbol{a}} \leftarrow F_{\mathrm{A}}(k_1, k_2, \cdots, k_8); \quad v \leftarrow (k_5 - k_8 - k_7) \bmod 2;$$

$$\text{For}(i_1 \leftarrow 0; i_1 \leq k_5 + k_6 + v; i_1 \leftarrow i_1 + 1)\{$$

$$\text{For}(i_2 \leftarrow 0; i_2 \leq k_5 + k_6 + v; i_2 \leftarrow i_2 + 1)\{$$

$$\text{If}\big(a_{i_1,i_2} \neq 0\big)\{\ \ p \leftarrow i_1 \bmod 2;$$

$$\text{For}(i_3 \leftarrow 0; i_3 \leq p; i_3 \leftarrow i_3 + 1)\{\ \text{For}\left(i_4 \leftarrow 0; i_4 \leq \left[\frac{i_1 - p}{2}\right]; i_4 \leftarrow i_4 + 1\right)\{$$

$$\text{For}(i_5 \leftarrow 0; i_5 \leq i_2 + 2i_3 + 2i_4; i_5 \leftarrow i_5 + 1)\{$$

$$\text{For}(i_6 \leftarrow 0; i_6 \leq i_2 + 2i_3 + 2i_4 - i_5; i_6 \leftarrow i_6 + 1)\{$$

$$x \leftarrow \frac{(-1)^{i_4+i_5}(2i_3)!\left(\left[\frac{i_1-p}{2}\right]\right)!\,(i_2 + 2i_3 + 2i_4)!}{2^{i_2+4i_3+2i_4}(1 - 2i_3)(i_3!)^2 i_4!\, i_5!\, i_6!\left(\left[\frac{i_1-p}{2}\right] - i_4\right)!\,(i_2 + 2i_3 + 2i_4 - i_5 - i_6)!};$$

$$i \leftarrow k_9 + i_2 + 2i_3 + 2i_4 - 2i_5 - 2i_6; \quad j \leftarrow k_{10} - i_2 - 2i_3 - 2i_4 + 2i_6;$$

$$\text{For}\left(i_7 \leftarrow 0; i_7 \leq \left[\frac{k_{11} + 2i_5 + 1}{2}\right]; i_7 \leftarrow i_7 + 1\right)\{$$

$$y \leftarrow \frac{(k_{11} + 2i_5 + 1)!}{(k_{11} + 2i_5 + 1 - 2i_7)!\,(2i_7 + 1)!};$$

$$i_0 \leftarrow i + k_{11} + 2i_5 - 2i_7 + 2; \ \ j_0 \leftarrow j + 2i_7 + 2;$$

$$X \leftarrow X + xya_{i_1,i_2} \cdot F_{\mathrm{R(II)}}(i_0, j_0, \xi_1, \xi_2);$$

$$i_0 \leftarrow j + k_{11} + 2i_5 - 2i_7 + 2; \ \ j_0 \leftarrow i + 2i_7 + 2;$$

$$X \leftarrow X + xya_{i_1,i_2} \cdot F_{\mathrm{R(II)}}(i_0, j_0, \xi_2, \xi_1);$$

$$\}\}\}\}\}\}\}\}\quad \text{Return } X \leftarrow 2X;\}\}$$ //End.

### *6. Subfunctions in repulsive energy*

Input: $\left(n_1, l_1, m_1, I_1, \xi_1, n_2, l_2, m_2, I_2, \xi_2, \vec{\boldsymbol{k}}\right)$;

Output: $X = \left\langle \varphi_{[\![1]\!]}^{*(0)}(\vec{\boldsymbol{r}}_1)\varphi_{[\![2]\!]}^{*(0)}(\vec{\boldsymbol{r}}_2) \middle| \frac{Y^{*(2)}(\vec{\boldsymbol{k}};\vec{\boldsymbol{\theta}},\vec{\boldsymbol{\phi}})r_1^{k_9}r_2^{k_{10}}}{r_{1,2}} \middle| \varphi_{[\![1]\!]}^{*(0)}(\vec{\boldsymbol{r}}_1)\varphi_{[\![2]\!]}^{*(0)}(\vec{\boldsymbol{r}}_2) \right\rangle$;

Algorithmic process:

$$F_\varphi\left(n_1, l_1, m_1, I_1, \xi_1, n_2, l_2, m_2, I_2, \xi_2, \vec{\boldsymbol{k}}\right)\{$$

$$X \leftarrow 0;$$

$$\text{If}(m_1 \geq 0)\{p_1 \leftarrow 0;\}\ \text{Else}\{p_1 \leftarrow 1;\}$$

$$\text{If}(m_2 \geq 0)\{p_2 \leftarrow 0;\}\ \text{Else}\{p_2 \leftarrow 1;\}$$

$$K_1 \leftarrow l_1 - |m_1| + 0.5 - \sqrt{0.25 - 2\delta_2(Z)};$$

$$K_2 \leftarrow l_2 - |m_2| + 0.5 - \sqrt{0.25 - 2\delta_2(Z)};$$

$$L_1 \leftarrow K_1 - (-1)^{I_1}\sqrt{m_1^2 - 2\delta_1(Z)};$$

$$L_2 \leftarrow K_2 - (-1)^{I_2}\sqrt{m_2^2 - 2\delta_1(Z)};$$

$$\text{For}\left(s_1 \leftarrow 0; s_1 \leq \left[\frac{|m_1| - p_1}{2}\right]; s_1 \leftarrow s_1 + 1\right)\{$$

$$\text{For}\left(s_2 \leftarrow 0; s_2 \leq \left[\frac{|m_1| - p_1}{2}\right]; s_2 \leftarrow s_2 + 1\right)\{$$

$$x_1 \leftarrow \frac{(-1)^{s_1+s_2}4^{|m_1|-s_1-s_2-1}(|m_1| - s_1 - 1)!\,(|m_1| - s_2 - 1)!}{(|m_1| - 2s_1 - p_1)!\,(|m_1| - 2s_2 - p_1)!\,s_1!\,s_2!};$$

$$\text{If}(m_1 > 0)\{x_1 \leftarrow m_1^2 x_1;\}$$

$$\text{For}\left(s_3 \leftarrow 0; s_3 \leq \left[\frac{|m_2| - p_2}{2}\right]; s_3 \leftarrow s_3 + 1\right)\{$$

$$\text{For}\left(s_4 \leftarrow 0; s_4 \leq \left[\frac{|m_2| - p_2}{2}\right]; s_4 \leftarrow s_4 + 1\right)\{$$

$$x_2 \leftarrow \frac{(-1)^{s_3+s_4}4^{|m_2|-s_3-s_4-1}(|m_2| - s_3 - 1)!\,(|m_2| - s_4 - 1)!}{(|m_2| - 2s_3 - p_2)!\,(|m_2| - 2s_4 - p_2)!\,s_3!\,s_4!}x_1;$$

$$\text{If}(m_2 > 0)\{x_2 \leftarrow m_2^2 x_2;\}$$

$$\text{For}\left(s_5 \leftarrow 0; s_5 \leq \left[\frac{l_1 - |m_1|}{2}\right]; s_5 \leftarrow s_5 + 1\right)\{$$

$$\text{If}(s_5 = 0)\left\{a_{1,0} \leftarrow 1;\right\}$$

$$\text{Else}\left\{a_{1,s_5} \leftarrow -\frac{(K_1 - 2s_5 + 2)(K_1 - 2s_5 + 1) + 2\delta_2(Z)}{2s_5(2L_1 - 2s_5 + 1)} a_{1,s_5-1};\right\}$$

$$\text{For}\left(s_6 \leftarrow 0; s_6 \leq \left[\frac{l_1 - |m_1|}{2}\right]; s_6 \leftarrow s_6 + 1\right)\{$$

$$\text{If}(s_6 = 0)\left\{a_{2,0} \leftarrow 1;\right\}$$

$$\text{Else}\left\{a_{2,s_6} \leftarrow -\frac{(K_1 - 2s_6 + 2)(K_1 - 2s_6 + 1) + 2\delta_2(Z)}{2s_6(2L_1 - 2s_6 + 1)} a_{2,s_6-1};\right\}$$

$$\text{For}\left(s_7 \leftarrow 0; s_7 \leq \left[\frac{l_2 - |m_2|}{2}\right]; s_7 \leftarrow s_7 + 1\right)\{$$

$$\text{If}(s_7 = 0)\left\{a_{3,0} \leftarrow 1;\right\}$$

$$\text{Else}\left\{a_{3,s_7} \leftarrow -\frac{(K_2 - 2s_7 + 2)(K_2 - 2s_7 + 1) + 2\delta_2(Z)}{2s_7(2L_2 - 2s_7 + 1)} a_{3,s_7-1};\right\}$$

$$\text{For}\left(s_8 \leftarrow 0; s_8 \leq \left[\frac{l_2 - |m_2|}{2}\right]; s_8 \leftarrow s_8 + 1\right)\{$$

$$\text{If}(s_7 = 0)\left\{a_{4,0} \leftarrow 1;\right\}$$

$$\text{Else}\left\{a_{4,s_8} \leftarrow -\frac{(K_2 - 2s_8 + 2)(K_2 - 2s_8 + 1) + 2\delta_2(Z)}{2s_8(2L_2 - 2s_8 + 1)} a_{4,s_8-1};\right\}$$

$$x_3 \leftarrow a_{1,s_5} a_{2,s_6} a_{3,s_7} a_{4,s_8} x_2;$$

$$\text{For}(s_9 \leftarrow 0; s_9 \leq n_1 - l_1 - 1; s_9 \leftarrow s_9 + 1)\{$$

$$\text{If}(s_9 = 0)\left\{b_{1,0} \leftarrow 1;\right\}$$

$$\text{Else}\left\{b_{1,s_9} \leftarrow \frac{\left(2s_9 - 1 + \sqrt{(2L_1 + 1)^2 - 8\delta_3(Z)}\right)\xi_1 - 2Z - 2\delta_0(Z)}{s_9\left(s_9 + \sqrt{(2L_1 + 1)^2 - 8\delta_3(Z)}\right)} b_{1,s_9-1};\right\}$$

$$\text{For}(s_{10} \leftarrow 0; s_{10} \leq n_1 - l_1 - 1; s_{10} \leftarrow s_{10} + 1)\{$$

$$\text{If}(s_{10}=0)\{b_{2,0}\leftarrow 1;\}$$

$$\text{Else}\left\{b_{2,s_{10}}\leftarrow\frac{\left(2s_{10}-1+\sqrt{(2L_1+1)^2-8\delta_3(Z)}\right)\xi_1-2Z-2\delta_0(Z)}{s_{10}\left(s_{10}+\sqrt{(2L_1+1)^2-8\delta_3(Z)}\right)}b_{2,s_{10}-1};\right\}$$

$$\text{For}(s_{11}\leftarrow 0;s_{11}\le n_2-l_2-1;s_{11}\leftarrow s_{11}+1)\{$$

$$\text{If}(s_{11}=0)\{b_{3,0}\leftarrow 1;\}$$

$$\text{Else}\left\{b_{3,s_{11}}\leftarrow\frac{\left(2s_{11}-1+\sqrt{(2L_2+1)^2-8\delta_3(Z)}\right)\xi_2-2Z-2\delta_0(Z)}{s_{11}\left(s_{11}+\sqrt{(2L_2+1)^2-8\delta_3(Z)}\right)}b_{3,s_{11}-1};\right\}$$

$$\text{For}(s_{12}\leftarrow 0;s_{12}\le n_2-l_2-1;s_{12}\leftarrow s_{12}+1)\{$$

$$\text{If}(s_{12}=0)\{b_{4,0}\leftarrow 1;\}$$

$$\text{Else}\left\{b_{4,s_{12}}\leftarrow\frac{\left(2s_{12}-1+\sqrt{(2L_2+1)^2-8\delta_3(Z)}\right)\xi_2-2Z-2\delta_0(Z)}{s_{12}\left(s_{12}+\sqrt{(2L_2+1)^2-8\delta_3(Z)}\right)}b_{4,s_{12}-1};\right\}$$

$$x_4\leftarrow b_{1,s_9}b_{2,s_{10}}b_{3,s_{11}}b_{4,s_{12}}x_3;$$

$$w_1\leftarrow k_1+2|m_1|;\ w_2\leftarrow k_2+2(l_1-|m_1|-s_5-s_6);w_3\leftarrow k_3+2p_1;$$

$$w_5\leftarrow k_5+2|m_2|;\ w_6\leftarrow k_6+2(l_2-|m_2|-s_7-s_8);w_7\leftarrow k_7+2p_2;$$

$$w_4\leftarrow k_4+2(|m_1|-p_1-s_1-s_2);\ w_8\leftarrow k_8+2(|m_2|-p_2-s_3-s_4);$$

$$w_9\leftarrow k_9+2l_1+s_9+s_{10};\ w_{10}\leftarrow k_{10}+2l_2+s_{11}+s_{12};w_{11}\leftarrow -1;$$

$$X\leftarrow X+x_4F_{ee}(\vec{\boldsymbol{w}},2\xi_1,2\xi_2);$$

$$\}\}\}\}\}\}\}\}\}\}\}\}\ \text{Return}\ X;\}$$ //End.

## V. Conclusion

In the ab initio calculation of the energy level fine structure of atoms and molecules, the two-step method in relativistic quantum mechanics can be replaced by the solution of a Schrödinger-like equation, and the lack of singlets (missing roots) of Hartree-Fock method without spin function is only a mathematical calculation problem, which cannot be a natural conclusion of the existence of spin function, and can be proved by its numerical experiments, such as the calculation of hydrogen-like atoms and the HAV of other atoms, omitted.

**Funding:** No funding.

**Competing interests:** I declare have no competing interests.

**Author Contributions：** All authors contribute equally.